\documentclass[twoside]{article}
\usepackage{fleqn,espcrc2}

\usepackage{epsfig}
\usepackage{graphicx}
\usepackage[figuresright]{rotating}

\newcommand{\be}{\begin{equation}}
\newcommand{\ee}{\end{equation}}
\newcommand{\ba}{\begin{eqnarray}}
\newcommand{\ea}{\end{eqnarray}}

\newcommand{\AmS}{{\protect\the\textfont2
  A\kern-.1667em\lower.5ex\hbox{M}\kern-.125emS}}
\def\spose#1{\hbox to 0pt{#1\hss}}
\def\ltapprox{\mathrel{\spose{\lower 3pt\hbox{$\mathchar"218$}}
 \raise 2.0pt\hbox{$\mathchar"13C$}}}

\def\p{^\prime}

\def\n{\noindent}

\title{A numerical study of Goldstone-mode effects and scaling functions 
of the three-dimensional $O(2)$ model
       \thanks{Talk given by J. Engels, the work was supported by DFG
        Grant No. Ka 1198/4-1} }

\author{J. Engels\address{Fakult\"at f\"ur Physik, Universit\"at Bielefeld, \\
        D-33615 Bielefeld, Germany},
        S. Holtmann\addressmark ,
        T. Mendes\addressmark , 
        and T. Schulze\addressmark}

\begin{document}

\begin{abstract}
We investigate numerically the three-dimensional $O(2)$ model on $8^3-160^3$ 
lattices as a function of the magnetic field $H$. In the low-temperature phase
we verify the $H$-dependence of the magnetization $M$ induced by the Goldstone 
modes and determine $M$ in the thermodynamic limit on the coexistence line 
both by extrapolation and by chiral perturbation theory. We compute two
critical amplitudes from the scaling behaviours on the coexistence line and on
the critical line. In both cases we find negative corrections to scaling.  
With additional high temperature data we calculate the scaling function and
show that it has a smaller slope than that of the $O(4)$ model. For future 
tests of QCD lattice data we study as well finite-size-scaling functions.
\vspace{1pc}
\end{abstract}

\maketitle

\section{INTRODUCTION}

$O(N)$ models are of general relevance to condensed matter 
physics and to quantum field theory, because many physical
systems exhibit a second-order phase transition with the same
universal properties.
Due to the existence of massless Goldstone modes  
in $O(N)$ models with $N>1$ and dimension $d=3$ and 4 \cite{madras} 
singularities are expected on the whole coexistence line $T<T_c, H=0$,   
in addition to the known critical behaviour at $T_c$. Recently these 
predictions have been confirmed by simulations of the
$3d$ $O(4)$ model \cite{EM}.
A further motivation for studying $3d$ $O(N)$ models is their relation 
to quantum chromodynamics (QCD). The QCD chiral phase transition 
for two light-quark flavors is supposed to be of second order in 
the continuum limit and to be in the same universality 
class as the $3d$ $O(4)$ model \cite{PW}. In the staggered formulation
of QCD on the lattice a part of chiral symmetry is remaining and that is 
$O(2)$. For the comparison to QCD lattice data it is therefore important
to know the $O(2)$ and $O(4)$ universal scaling functions. 

The $O(2)$-invariant nonlinear $\sigma$-model (or $XY$ model) 
on a $d-$dimensional hypercubic lattice is defined by
\be
\beta\,{\cal H}\;=\;-J \,\sum_{<i,j>} {\bf S}_i\cdot {\bf S}_j
         \;-\; {\bf H}\cdot\,\sum_{i} {\bf S}_i \;.
\ee
${\bf S}_i$ is an 2-component unit vector at site $i$
with a longitudinal (parallel to the magnetic 
field ${\bf H}$) and a transverse component 
\be
{\bf S}_i\; =\; S_i^{\parallel} {\bf \hat H} + {\bf S}_i^{\perp} ~.
\ee
The order parameter of the system, the magnetization $M$, is given by
\be
M \;=\; <\!\frac{1}{V}\sum_{i} S_i^{\parallel}>\; =\; <  S^{\parallel}>~.
\ee
There is a longitudinal and a transverse susceptibility
\ba
\chi_L\!\! &\!=\!&\!\! {\partial M \over \partial H}
 \;=\; V(<{ S^{\parallel}}^2>-M^2)~, \label{chil}\\
\chi_T\!\! &\!=\!&\!\! V < {{\bf S}^{\perp}}^2>  \;=\;{M \over H}
~. \label{chit}
\ea
In the broken phase ($T<T_c$) the magnetization attains a finite value 
$M(T,0)$ at $H=0$. Consequently the transverse susceptibility diverges
as $H^{-1}$ when $H\to0$ for all $T<T_c$. It is non-trivial that 
also the longitudinal susceptibility is diverging on the coexistence 
line for $2<d\leq4$. The predicted \cite{WZ} divergence for $d=3$ is 
\be
\chi_L(T<T_c,H)\;\sim\; H^{-1/2}~,
\label{chiL}
\ee
which is equivalent to an $H^{1/2}$-behaviour of the magnetization near 
the coexistence curve
\be
M(T<T_c,H)\;=\;M(T,0)\,+\,c\,H^{1/2}~.
\label{magn}
\ee
In finite volumes and $H\rightarrow 0$ the Goldstone modes induce 
strong finite-size effects at all $T<T_c$.


\section{NUMERICAL RESULTS}
\label{section:results}

Our simulations were done on three-di\-men\-sio\-nal lattices with periodic 
boundary conditions and linear extensions $L=8-160$ using the cluster 
algorithm. In order to eliminate finite-size effects
we simulated for increasingly larger values of $L$ 
at fixed values of $J=1/T$ (i.e.\ fixed temperature $T$) and $H$. For $1/T_c$ 
we took the value $J_c = 0.454165(4)$ from Ref. \cite{Spain}.

\begin{figure}[hb!]
\begin{center}
   \epsfig{bbllx=68,bblly=307,bburx=507,bbury=534,
        file=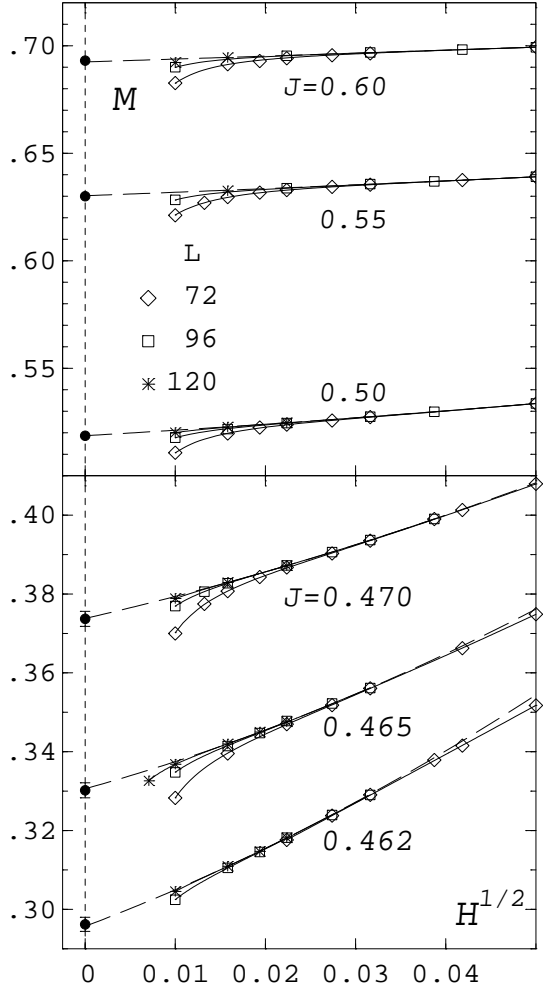,height=63mm,angle=-90}
\end{center}
\caption{The magnetization vs. $H^{1/2}$ in the 
low-temperature region for fixed $J$ and various $L$.} 
\label{fig:mroot}
\end{figure}

In Fig.\ \ref{fig:mroot} we show the data for the magnetization as a 
function of $H^{1/2}$ for six fixed values of $J$ in the low-temperature 
phase.  The picture is rather similar to the one obtained in $O(4)$ 
\cite{EM}: strong finite-size effects appear for small $H$ and persist 
as one moves away from $T_c$, the results from the largest lattices are 
at first sight linear in $H^{1/2}$, as predicted by Eq.\ \ref{magn}. 
Very close to $H=0$ the fixed temperature curves become 
slightly flatter, leading to a higher value for $M(T,0)$ than expected 
from the data at larger $H$ values. This behaviour is more pronounced 
close to $T_c$ than at lower temperatures. In order to extrapolate the data to 
$H\rightarrow 0$ and $V\rightarrow \infty$ we apply two different 
strategies. The first is to extend the linear form in $H^{1/2}$, 
Eq.\ \ref{magn}, to a quadratic one 
\be
M(T<T_c,H)\,=\,M(T,0)\,+\,c_1\,H^{1/2}+\,c_2\,H ,
\label{magn2}
\ee
and to fit the data from the largest lattices, which we assume to represent
data on an infinite volume lattice, to this form. The second way to find
$M(T,0)$ is just opposite to the first. Here we exploit the $L$ or volume
dependence at fixed $J$ and fixed small $H$ to determine via chiral 
perturbation theory (CPT) \cite{HL} the magnetization $\Sigma$ of the continuum 
theory for $V\rightarrow \infty,~H=0$, which is related to 
$M(T,0)$ by 
\be
M(T,0)\;=\;{\Sigma \over \sqrt{J}}~.
\label{magn3}
\ee
We observe in Fig.\ \ref{fig:mroot} a remarkable coincidence of
\newpage 

\n the fits 
according to Eq.\ \ref{magn2} (dashed lines) with the CPT results at 
$H=0$ (filled circles). In the neighbourhood of $T_c$ the results for 
$M(T,0)$ should show the usual critical behaviour. Since we 
expect here sizeable corrections to scaling \cite{Hase} we make the 
following ansatz to determine the critical amplitude $B$ (${\bar t}=T_c-T$)  
\be
M(T\ltapprox T_c,0)=B{\bar t}^{\beta}[1+
                    b_1{\bar t}^{\omega \nu}+b_2{\bar t}]~.
\label{cocrit}
\ee
\n Here and in the following we use the critical exponents 
\begin{figure}[ht!]
   \epsfig{bbllx=97,bblly=264,bburx=421,bbury=587,
       file=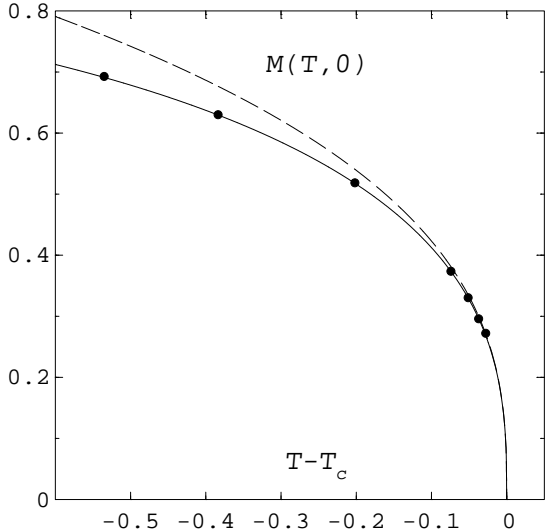, width=65mm}
\caption{The magnetization at $H=0$ vs. $T-T_c$ with the fit\ (\ref{cocrit}) 
(solid line) and its leading part (dashed line).}
\label{fig:mcoex}
\end{figure}
from Ref.\ \cite{Hase} 
\be
\beta=0.3490(6),~ \nu=0.6723(11),~ \omega=0.79(2)~.
\label{Hasex}
\ee
A fit to all extrapolated points gives
$$
B=0.945(5),~\!b_1=-0.053(23),~\! b_2=-0.098(23) .
$$
In Fig.\ \ref{fig:mcoex} we show this fit and also the leading term
separately. As the critical point is reached the $H$-dependence of 
the magnetization changes to satisfy critical scaling. We therefore fit 
the data from the largest lattice sizes at $T_c$ to the form
\be
M(T_c,H) \;=\;d_c H^{1/\delta}[ 1 + d_c^1 H^{\omega \nu_c}]~.
\label{hcrit}
\ee
A further term proportional to $H$ is unnecessary here, because the corrections
to scaling are much smaller than on the coexistence line. The largest 
$L$ data can be fitted very well with the ansatz (\ref{hcrit}) and the 
critical exponents as input. We find $d_c=0.978(2)$ and $d_c^1=-0.075(5)$, 
that is again negative corrections to scaling.

\section{THE SCALING FUNCTION}
\label{section:sca_fun}

In the thermodynamic limit, the dependence of the magnetization on
temperature and magnetic field can be expressed \cite{EM} in the form 
\newpage 
\begin{figure}[ht!]
\begin{center}
   \epsfig{bbllx=127,bblly=279,bburx=451,bbury=587,
        file=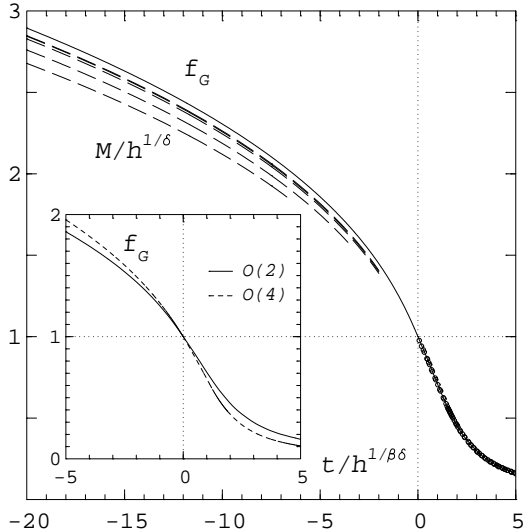,width=65mm}
\end{center}
\caption{The scaling function $f_G$ (solid line) and results for 
$M/h^{1/\delta}$ at $J=0.55,0.50,0.47,0.465$ and 0.462 (dashed lines), 
starting with the lowest curve. The circles are single data points,
the inset shows $f_G$ for $O(2)$ and $O(4)$.  } 
\label{fig:fgscale}
\end{figure}
\be
M\;=\;h^{1/\delta} f_G(t/h^{1/\beta\delta})~,
\label{ftous}
\ee
where $f_G$ is a universal scaling function and $t$ and $h$ are the normalized 
reduced temperature $t=(T-T_c)/T_0$ and magnetic field $h=H/H_0$. Here 
\be
H_0 = d_c^{-\delta} = 1.11(1)~,~T_0 = B^{-1/\beta} = 1.18(2)~, 
\ee 
which implies $f_G(0)=1$ and $f_G(t<0,h\rightarrow 0)\rightarrow (-t)^{\beta}
h^{-1/\delta}$. Obviously, $f_G$ does not account for possible corrections to
scaling and is the leading term in the Taylor expansion in $h^{\omega\nu_c}$
of a more general form
\be 
Mh^{-1/\delta}\; =\; \Psi (th^{-1/\beta\delta}, h^{\omega\nu_c})~.
\ee
We therefore perform quadratic fits to our data in $h^{\omega\nu_c}$ 
at fixed values of $th^{-1/\beta\delta}$ in the low-temperature region,
where the corrections are strong. In the high-temperature region the data 
scale directly. In Fig.\ \ref{fig:fgscale} we have plotted $Mh^{-1/\delta}$ 
from data with $H \leq 0.0075$ and $0.43 \le J \le 0.55$ and the final 
result for the scaling function $f_G$. An alternative scaling form is that
of Widom and Griffiths. It is discussed in Ref. \cite{O2}, where more 
\newpage
\begin{figure}[ht!]
\begin{center}
   \epsfig{bbllx=93,bblly=256,bburx=497,bbury=580,
        file=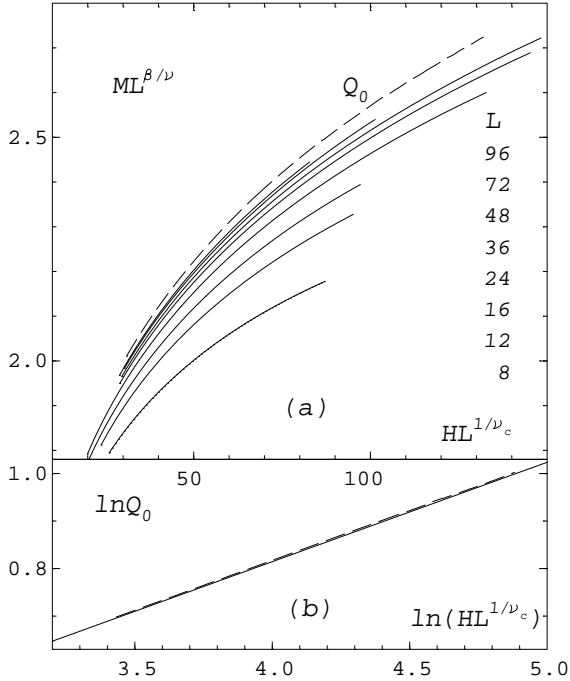,width=82mm,angle=-90}
\end{center}
\caption{ (a) $ML^{\beta/\nu}$ from reweighted data from lattices with 
different $L$ (solid lines) and the scaling function $Q_0$ (dashed line)
at $z=0$. (b) Comparison of $\ln Q_0$ to its asymptotic value (line).} 
\label{fig:fssq}
\end{figure}
\n details of our calculations can be found. 

In Ref.\ \cite{MILC} staggered lattice QCD data
for $N_{\tau}=4$ were compared to the $O(4)$ scaling function. The test failed
because the data were indicating a steeper scaling function. Since
the $O(2)$ scaling function is even flatter than the one for $O(4)$, as 
can be seen from the inset of Fig.\ \ref{fig:fgscale}, the situation will be
worse there. A way out may be the comparison to finite-size-scaling functions,
since lattice QCD is presumably far from the thermodynamic limit.

\subsection{Finite-size-scaling functions}

\medskip
The general form of the finite-size-scaling function for the 
magnetization is given by
\be
M = L^{-\beta/\nu} \Phi(tL^{1/\nu},hL^{1/\nu_c},L^{-\omega})~.
\ee
On lines of fixed $z=th^{-1/\beta\delta}$ and after expanding in $L^{-\omega}$
we have
\newpage
\be
M = L^{-\beta/\nu} Q_{0z}(hL^{1/\nu_c})+\dots~,
\ee
where $Q_{0z}$ is a {\em universal} function. Examples of lines of fixed $z$
are the critical line where $z=0$ and the pseudocritical line, the line of
maximum positions of the susceptibility $\chi_L$ in the $(t,h)$-plane
for $V\rightarrow \infty$. The scaling function contains also all 
information about $\chi_L$, because 
\be
\chi_L={\partial M\over \partial H}={h^{1/\delta-1} \over H_0 \delta}
\left ( f_G(z) - {z\over \beta}f_G\p(z) \right)~.
\label{max}
\ee
Evidently, the maximum of $\chi_L$ at fixed $h$ and varying $t$
is at the maximum point $z_p$ of the function in the brackets
of Eq. (\ref{max}). $z_p$ is again a universal quantity and we find
$z_p=1.556\pm 0.10$ from (\ref{max}). As a check we have also determined the
peak positions on lattices with $L=24-96$ which extrapolate to  
$z_p=1.65\pm 0.10$ for $L\rightarrow \infty$.
In Fig.\ \ref{fig:fssq}a we show $ML^{\beta/\nu}$ from the reweighted data
on the critical line for various $L$ values and also the scaling function
$Q_0(z=0)$. For  $L\rightarrow \infty$ the scaling function $Q_{0z}$ is 
related to $f_G$ by
\be
Q_{0z}\rightarrow f_G(z)(hL^{1/\nu_c})^{1/\delta}~.
\ee
From Fig.\ \ref{fig:fssq}b we see that $Q_0(z=0)$ seems to be asymptotic 
in the whole variable range considered. At present we investigate the case  
$z=z_p$.

\end{document}